\documentclass[journal]{IEEEtran}
\usepackage{color}
\usepackage{cite}
\usepackage{graphicx}
\usepackage{amsmath}
\usepackage{multicol}

\input{psfig}
\usepackage{amssymb}

\begin{document}
\title{Noise Invalidation Denoising}
\author{Soosan Beheshti,~\IEEEmembership{Senior Member,~IEEE,} Masoud Hashemi,~\IEEEmembership{Student
Member,~IEEE,} and Xiao-Ping Zhang,~\IEEEmembership{Senior
Member,~IEEE, Nima Nikvand,~\IEEEmembership{Student
Member,~IEEE,}}
\thanks{*S. Beheshti, corresponding author, (soosan@ee.ryerson.ca, phone: 416 979 5000
ex:4906), Masoud Hashemi (s3hashem@ee.ryerson.ca), Xiao-Ping Zhang
(xzhang@ee.ryerson.ca)) are with the Department of Electrical and
Computer Engineering, Ryerson University, Toronto, ON, M5B 2K3,
Canada. Nima Nivand(nnikvand@uwaterloo.ca) is with the Department of Electrical and
Computer Engineering, University of Waterloo, Waterloo, ON, N2L 3G1, Canada.}} \maketitle
\begin{abstract}
A denoising technique based on noise invalidation is proposed. 
The adaptive approach derives a noise signature from the noise
order statistics and utilizes the signature to denoise the data.
The novelty of this approach is in presenting a general-purpose
denoising in the sense that it does not need to employ any
particular assumption on the structure of the noise-free signal,
such as data smoothness or sparsity of the coefficients. An
advantage of the method is in denoising the corrupted data in any
complete basis transformation (orthogonal or non-orthogonal).
Experimental results show that the proposed method, called Noise
Invalidation Denoising (NIDe), outperforms 
existing denoising approaches in terms of Mean Square Error (MSE).
\end{abstract}
\section{Introduction}
\label{sec:intro}
 Data denoising approaches are well known methods with
  presence in diverse applications and have been studied and
 developed in research areas ranging from communications to biomedical signal analysis.
 In denoising techniques, multi-resolution representations of the data is generally used. For example, wavelet shrinkage is based on
rejecting those wavelet coefficients that are smaller than a
certain value and keeping the remaining coefficients. Thus, the
problem of removing noise from a set of observed data is
transformed into finding a proper threshold for the
 data coefficients. The pioneer shrinkage methods, such as VisuShrink and SureShrink, propose
thresholds that are functions of the noise variance and the data
length \cite{visu,sure}. Over the past fifteen years, several
thresholding approaches such as
\cite{Blu,XPZhang2,Chang,Chang2,Abramovitch} have been developed.
These methods provide optimum thresholds by focusing on certain
properties of the noise-free signal, and they are proposed for
particular applications, mostly for the purpose of image
denoising.
Unlike these approaches, the method presented in this paper
focuses only on the properties of the additive noise. By relying
on the noise statistics, the method defines a probabilistic region
of confidence for the noise coefficients. Consequently, it
validates those observed coefficients that are out of the noise
confidence region and contain noiseless dominant
 parts. In the presence of additive colored noise, the required data preprocessing and/or whitening filters may damage the desired noiseless data. Dealing with the colored noise without whitening is especially critical if the noise correlation is due to the non-orthogonality of the basis. The proposed method denoises such data by using only the colored noise statistics in the invalidation step. The principle behind the proposed approach is simple yet
powerful as it takes advantage of the properties for the additive
noise and demonstrates efficiency in applications for
general-purpose denoising.

The paper is organized as follows. In Section \ref{sec:Problem
Formulation} the considered problem is formulated and revisits the
philosophy of thresholding. Section \ref{sec:Noise Statistics}
derives a noise signature from the noise statistics. Section
\ref{sec2} presents the proposed noise invalidation approach. In
Section \ref{sec:simulation} simulation results are provided, and
conclusions are drawn in Section \ref{con}.

{\em Notations:} We use capital letters $V$, $W$, and $\Theta$ for
random variables. Samples of these variables are shown with small
letters $v$, $w$, and $\theta$.
\section{Problem Formulation and Motivation}
\label{sec:Problem Formulation} Noise-free data vector of length
$N$, $\bar{y}^N= [\bar{y}_1,\cdots,\bar{y}_N]^T$ is corrupted by
an additive white Gaussian random process $w^N
=[w_1,\cdots,w_N]^T$ with zero mean and variance of $\sigma^2$.
The observed data $y^N=[y_1,\cdots,y_N]^T$ is
\begin{eqnarray}
y[n]=\bar y[n]+w[n]
\end{eqnarray}
 and can be
expressed in terms of a desired orthonormal basis such that \footnote{Inner product of real vectors $a$ and $b$ is denoted by $<a,b>=a^Tb$.}:
\begin{equation}
\theta[i] = <s_i,y^N>
\end{equation}
where $s_i$ is an element of orthonormal basis $S =
[s_1,s_2,\cdots,s_N]$ 
and the desired unavailable coefficient is
\begin{eqnarray}
 {\bar \theta}_i =
<s_i,\bar{y}^N>
\end{eqnarray}
Therefore, the following holds
\begin{equation}
\bar{y}^N = \sum_{i=1}^N \bar \theta_is_i, \ \ \ y^N =
\sum_{i=1}^N\theta_is_i
\end{equation}
and thanks to the orthonormality of the selected basis \textit{S},
the noise coefficients,
\begin{eqnarray}
v^N = [v_1,\cdots,v_N]^T=<S,w^N>
\end{eqnarray}
 in the available coefficients
 \begin{eqnarray}
\theta_i = \bar \theta_i+v_i, \label{coef}
\end{eqnarray}
 are also
independent identically distributed random variables with the same
mean and variance of the noise
\begin{equation}
E(V_i) = 0,\;var(V_i) = \sigma^2 \;\; 1\leq i\leq
N \label{V}
\end{equation}
The observed coefficients are soft thresholded by threshold $T_s$
\begin{equation}
\hat{\theta}_{T_s} =\left \{ \begin{array}{lr}
sgn(\theta[i])(|\theta[i]| - T_s) & \textit{if} \ \  |\theta[i]|
\geq
T_s\\
0 & \textit{Otherwise}\\
\end{array}
\right.
\end{equation}
The estimate of the noise-free data
 with this threshold is
\begin{equation}
\hat{y}^N_{T_S} = \sum_{i=1}^N \hat{\theta}_{T_s}[i] s_i
\label{sahand}
\end{equation}
We provide the optimum threshold by a noise invalidation process.
\subsection{Main Idea of Thresholding}\label{main}
Traditionally, the main idea in thresholding is to choose a value
for which all the coefficients with absolute values smaller than
that value are thrown away. The rejected coefficients are the ones
that we cannot fully determine their contribution to the
noise-free data and therefore, we decide to dismiss them.
Nevertheless, there is always a chance that some samples of the
pure noise are above the threshold and some noise-free
coefficients get buried among the discarded coefficients due to
the additive noise. The existing thresholding methods usually
focus on a class of signals, such as images, to provide a proper
threshold and evaluate the quality of the thresholds based on some
performance criterion such as the mean square error (MSE). Among
the thresholding methods, the ones with a more general assumption
on the noise-free data are VisuShrink and SureShrink, which rely
on some form of piecewise smoothness of the wavelet coefficients.
The rest of the thresholding methods are application oriented, for
example the BayesShrink or other image denoising methods that use
properties such as generalized Gaussian distribution (GGD) for the
noise-free image itself. Another example is the adaptive denoising
approach in \cite{XPZhang2} for neural network applications.

Here, we present a general-purpose thresholding approach
that does not need to exploit any particular property of the noise-free
data itself. Consequently, the thresholding approach should be
invariant with respect to the order of the data. So if the data is
reordered and put in an ascending order based on its absolute
value, the optimum threshold should remain the same.  In this
case, choosing the $m$th coefficient of the ordered data as the
 threshold is equivalent to throwing away the first $m-1$ coefficients.

For a general-purpose threshold, instead of concentrating on the
properties of the remaining coefficients after thresholding, it is
logical to focus on the dismissed coefficients. These coefficients
are discarded because they are  attributed to the noise or very
noisy coefficients. It is rational to equivalently state that
these coefficients are discarded since they behave similarly to a
set of coefficients that can be generated by an associated
Gaussian distribution of the additive noise. In the following
section we present one of the signatures of a set that is
generated by this Gaussian distribution \footnote{Probabilistic
approach for finding a significant noise-free component in a
coefficients is discussed in a data denoising approach in
\cite{Pizurica}. However, in this
approach a Laplacian prior for a noise-free data is assumed. \\Preliminary work related to the proposed method is presented in \cite{ANIST}}. 

{\em Additive Noise Variance:} Thresholding methods, such as
SureShrink and VisuShrink, rely heavily  on the value of the
additive noise variance $\sigma^2$. In most practical applications
this value is not known. The estimate of the standard deviation is
usually provided by MAD approach where $\hat \sigma={\rm
MAD}/.6745$, and where MAD is the median of absolute value of
normalized fine scale wavelet coefficients \cite{sure}. The method
provided in this paper also requires the estimate of noise
variance, and we use the MAD method for this estimation. 
\section{Additive Noise
Signature}\label{sec:Noise Statistics}
Consider the additive noise random variable $V$ in (\ref{V}) with zero mean and finite variance. Define the signature function for any value $z$ and $v$ as $g(z,v)$ such that the mean and variance of this function over $V$ are finite values:
\begin{eqnarray}
E(g(z,V))&=&G_E(z) \label{onne2}\\
var(g(z,V))&=&G_{var}(z)\label{for4}
\end{eqnarray}
The signature for samples of a random process of length $N$, $V^N=[V_1,\cdots,V_N]^T$, with IID members that have the same distribution of $V$ is defined as
\begin{eqnarray}
g(z,v^N)=\frac{1}{N}\sum_{i=1}^Ng(z,v_i)\label{ave}
\end{eqnarray}
It follows that the expected value and variance of the signature are
\begin{eqnarray}
E(g(z,V^N))&=&G_E(z) \label{o2}\\
var(g(z,V^N))&=&\frac{1}{N}G_{var}(z)\label{f4}
\end{eqnarray}
Detailed are shown in Appendix \ref{mv}. For a large data length, while the mean is a finite fixed value, the variance becomes smaller. The use of such signatures in invalidation of the additive noise is explored with the following example.
\subsection{Signature Example: Absolute Noise Sorting (ANS)}
Consider a noise signature with the following form
\begin{equation}
g(z,v_i) =\left \{ \begin{array}{lr}
1 & \textit{if} \ \  |v_i| \leq z
\\
0 & |v_i| > z\\
\end{array}
\right.
\end{equation}
In this case, for the signature of the IID random process $V^N$ in (\ref{ave}) we have \cite{order,CDF1,CDF2}:
\begin{eqnarray*}
E(g(z,V^N))&=&F(z) \label{avenoise}\\
var(g(z,V^N))&=&\frac{1}{N}F(z)(1-F(z))
\end{eqnarray*}
where $F(\cdot)$ is the cdf of {\em absolute value} of the additive noise
\begin{eqnarray}
F(z)=2\phi(\frac{z}{\sigma})-1
\end{eqnarray}
 Details are provided in Appendix \ref{gol}. $\diamondsuit$

\begin{figure}[ht]
\centerline{\psfig{file=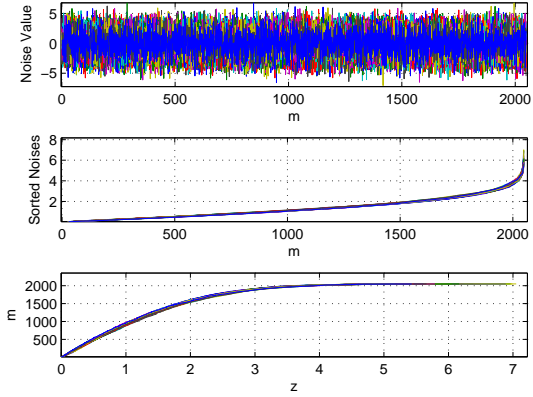,height=6cm}}
\caption{ Top figure: 100 runs of a zero mean Gaussian
distribution with unit variance and length 2048. Middle: The same
100 runs of the above figure sorted based on their absolute
values. Bottom: This is the middle figure with its vertical and
horizontal axes swapped ($m=Ng(z,v^N)$). } \label{gauss}
\end{figure}

 For each $z$, the value of $g(z,v^N)=m/N$ where $m$ is the number of samples of $v^N$ with absolute values smaller than $z$. Equivalently, when sorting $v^N$, the $m$th value is the largest $v_i$ that is smaller than $z$.
 Therefore, when sorting the $v_i$s, the index is $Ng(z,v^N)=m$.
 Figure \ref{gauss} illustrates the effect of sorting  and the role of the small variance in providing a noise signature.
The figure shows the behavior of 100 samples of Gaussian noise with unit variance
and length 2048. As the top figure shows, with a very high
probability, the values of this data are bounded between $\pm
3\sigma$.
However, if we sort the same data based on its absolute value in
the middle figure, the values collapse in a much denser area. Such behavior can be explained by the ANS signature as follows. The bottom figure show the result of swapping the horizontal and vertical axes of the middle figure. Here the horizontal axis is $z$ and the vertical shows 100 samples of $Ng(z,v^{N})$ where $N=2048$. As it is expected, these values are around mean $NF(z)$ with variance $F(z)(1-F(z))$. This will allow us
to define proper confidence regions, with a high probability $p$,
around the noise signature. Due to the noise signature structure, these
regions are considerably smaller than the corresponding confidence
regions of the Gaussian distribution of the additive noise itself.
\begin{figure}[ht!]
\centerline{\psfig{file=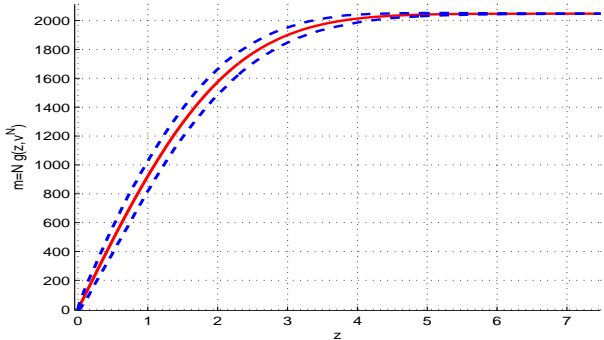,width=9cm,height=5cm}}
\caption{Solid line: Mean of the noise $g(z,V^N)$. Dashed lines
are upper and lower bounds with confidence probability 0.999997.}
\label{pnistmean2}
\end{figure}
Therefore, for each $z$ and for a high confidence probability $p$,
we can find $L_N(z)$ and $U_N(z)$ around the mean value of $F(z)$
such that
\begin{eqnarray}
Pr\{L_N(z) \leq g(z,v_N) \leq U_N(z) \}=p \label{bounds0}
\end{eqnarray}
For example, Figure \ref{pnistmean2} shows the bounds on
$g(z,V^N)$ for confidence probability p=0.999997 and with
$\sigma=2.5$.
\subsection{Confidence Region and Gaussian Estimate}
While it is straightforward to make a table of values of the
boundaries shown in Figure \ref{pnistmean2}, it is  possible to
use Gaussian estimates for the distributions of $g(z,v_N)$
 for large enough values of $N$ as $g(z,v_N)$ in (\ref{ave}) is average of
 $N$ independent variables. Using the Central Limit Theorem for this distribution, we have
  \footnote{The error
function is
\begin{eqnarray}
erf(x)=\frac{1}{\sqrt{\pi}}\int_0^{x} e^{-t^2} dt
\end{eqnarray}}
\begin{eqnarray}
Pr\{ \frac{|g(z,v_N) -F(z)|}{\lambda\sqrt{\frac{1}{N}F(z)(1-F(z))}}\leq 1\} \approx
erf(\frac{\lambda}{\sqrt{2}}) \label{eqn:erf}
\end{eqnarray}
 This estimates the boundaries in
(\ref{bounds0}) to be
\begin{eqnarray}
L_N(z)=F(z)-\lambda \sqrt{\frac{1}{N}F(z)(1-F(z))} \label{bounds}\\
U_N(z)=F(z)+\lambda \sqrt{\frac{1}{N}F(z)(1-F(z))}
\end{eqnarray}
 The choice
of $\lambda$ should be such that the probability is close to one
and at the same time the boundary is not very loose. In statistics
the three-sigma rule, or empirical rule, states that for a normal
distribution, almost all values lie within three standard
deviations of the mean\footnote{The following is known for
$\lambda=3, 4.5$ and $5$
\begin{eqnarray}
erf(3/\sqrt{2})=0.997300203937,\;\;\;\; erf(4.5/\sqrt{2})=0.999997
\end{eqnarray}}.
For a better quality measure, the six sigma approach increases the
standard deviation to 4.5 (equivalently $p=0.999997$).
Consequently, we suggest choosing $\lambda$ such that
$3\leq\lambda\leq 5$.  Interestingly, our experimental observation
shows that the threshold associated with $\lambda=4.5$ provides
the optimum threshold with respect to MSE in 90$\%$ of cases.
\section{Noise invalidation with Absolute Coefficient Sorting (ACS)}\label{sec2}
 The coefficients of our observed data is in form of $\theta_i=v_i+\bar \theta_i$ which has the same structure as the noise except its mean which is the noiseless coefficient $\bar \theta_i$. In this case
 \begin{eqnarray}
E(g(z,\Theta_i))=Pr(|v_i+\bar\theta_i| \leq z)=H(z,\bar\theta_i) \label{rahman}\\
var(g(z,\Theta_i))=H(z,\bar\theta_i)(1-H(z,\bar\theta_i)) \label{rahim}
\end{eqnarray}
 where
  \begin{eqnarray}
  H(z,\bar\theta_i)=\phi(\frac{z-\bar\theta_i}{\sigma})+\phi(\frac{z+\bar\theta_i}{\sigma})-1 \label{H}
\end{eqnarray}
Details are provided  in Appendix \ref{golnar}. $\diamondsuit$

\begin{figure}[ht!]
\centerline{\psfig{file=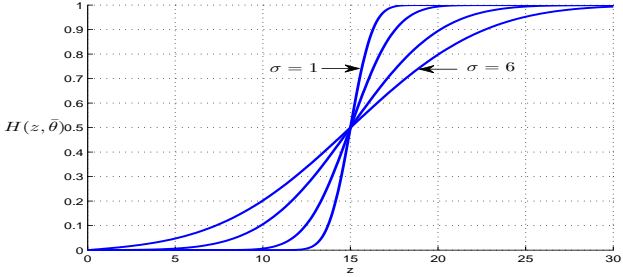,width=9cm,height=4cm}}
\caption{Expected value of $g(z,\Theta)$ when $\bar\theta=15$ and
the noise standard deviations $\sigma=1,2,4$ and $6$.}
\label{hist2}
\end{figure}

\begin{figure}[ht!]
\centerline{\psfig{file=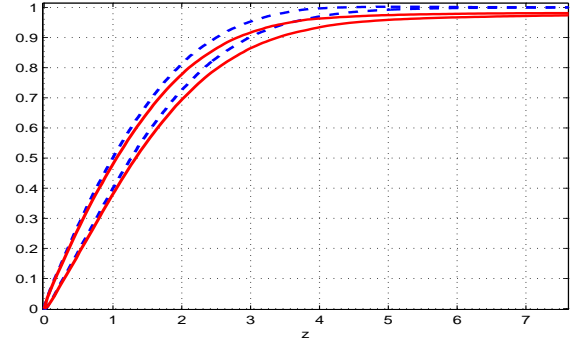,width=9cm,height=5cm}}
\caption{The area between the solid lines is the confidence region
of sorted absolute values of the noisy data coefficients of Blocks
signal (SNR=5) with probability 0.999997. The area between the
dashed lines is the noise confidence region with probability
0.999997.} \label{th}
\end{figure}

Figures \ref{hist1} and \ref{hist2} show typical behaviors of $H(z,\bar\theta_i)$ for various $\bar\theta_i$s and change of noise variance. 
\begin{figure}[ht!]
\centerline{\psfig{file=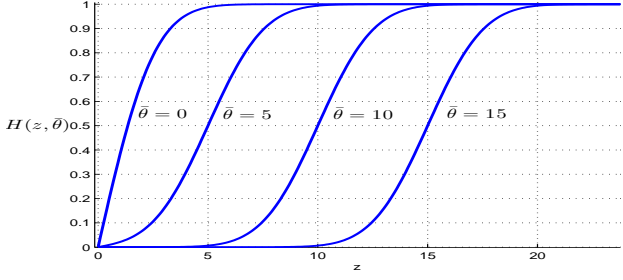,width=9cm,height=4cm}}
\caption{Expected value of $g(z,\Theta)$ for various values of
$\bar \theta$ when the additive noise variance is $\sigma=4$.}
\label{hist1}
\end{figure}
Sorting the coefficients in this case is analogous to calculation of
\begin{eqnarray}
g(z,\theta^N)=\frac{1}{N}\sum_{i=1}^Ng(z,\theta_i) \label{ave2}
\end{eqnarray}
which according to (\ref{rahman}) and (\ref{rahim}) has the following mean and variance
\begin{eqnarray}
Eg(z,\Theta^N)&=&\frac{1}{N}\sum_{i=1}^N H(z,\bar\theta_i) \label{avetheta}\\
var(g(z,\Theta^N))&=&\frac{1}{N^2}\sum_{i=1}^N
H(z,\bar\theta_i)(1-H(z,\bar\theta_i)) \label{vartheta}
\end{eqnarray}
Since the value of $H(z,\bar\theta_i)$ in (\ref{H}) is bounded between zero and one, the variance of this value is much less than its mean for large values of $N$. Therefore, a dense area will cover the sorted data with a high probability. This area becomes distinguished from the area covered by the sorted noise-only signal as the value of $z$ grows and as the nonzero coefficients become effective. This performance is illustrated in Figure \ref{th} which shows the area covered by the sorted noisy data for Blocks signal for when SNR=5. The figure also shows the behavior of the sorted noise-only data. As it can be seen with probability $0.99997$ there is no overlap between the sorted noise and sorted noisy data after a certain value of $z$.
\subsection{Noise Invalidation in Application}
\begin{figure}[ht!]
\centerline{\psfig{file=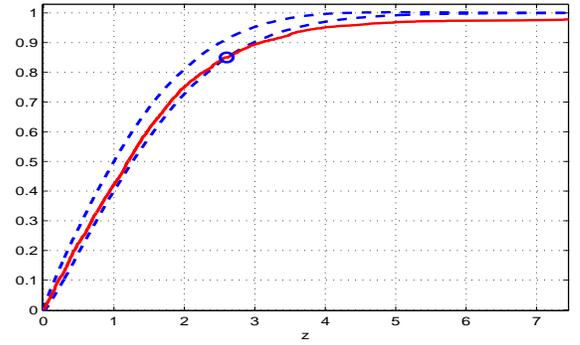,width=9cm,height=5cm}}
\caption{Solid line is the sorted absolute values of the observed
data coefficients crossing upper bound of the noise confidence
region at $z=2.6$ when the observed data is noisy Blocks signal
(SNR=5). The area between the dashed lines is the noise confidence
region with probability 0.999997.} \label{app}
\end{figure}

Using the noise sorting signature, it is possible to invalidate
the noisy coefficients with a high confidence. Figure \ref{app}
shows the application of the method. The confidence region for
the noise-only data is available upon knowing or estimating the noise
variance. As the sorted absolute noisy data leaves the noise
confidence region, it assures that the coefficients are becoming
more effective than the noisy part of the data. The largest $z$
value for which the departure occurs is the optimum threshold
($T^*$) for the noise validation problem
\begin{eqnarray}
T^*=max_z \forall z\leq x:  L_N(x)\nleqslant g(x,\theta^N)
\nleqslant U_N(x)
\end{eqnarray}

 \subsection{Colored Noise and Thresholding}
{\em Corollary:} If the additive noise is colored, the expected value of $g(z,V^N)$
and $g(z,\Theta^N)$ will remain the same as these expected values for
the white noise in (\ref{avenoise}) and (\ref{avetheta}).  For the
variance of the sorted noise we have
\begin{eqnarray}
var(g(z,V^N)))=\frac{1}{N}F(z)(1-F(z))+ \nonumber \\ \frac{1}{N^2}\sum_{i=1,j=1,i\neq j}^N cov(g(z,V_i)g(z,V_j))\\ \leq \frac{1}{N}F(z)(1-F(z))+ \nonumber \\ \frac{1}{N^2}\sum_{i=1,j=1,i\neq j}^N [F(\frac{z}{\sqrt{1+\rho_{ij}}})F(\frac{z}{\sqrt{1-\rho_{ij}}})-F^2(z)]
\end{eqnarray}
where
\begin{eqnarray}
\rho_{ij}=\frac{R_{vv}(i-j)}{R_{vv}(0)}
\end{eqnarray}
with $R_{vv}(0)=\sigma^2$.

{\em Proof:} In Appendix \ref{gol2}.

The variance for the sorted noisy
 data is also provided in Appendix \ref{gol2}. As the variance indicates, the wider
 is the autocorrelation of the noise process with itself, the wider
is the signature region of the noise and noisy data and therefore,
as it is expected, it may become more difficult to distinguish the data from the
noise.
\section{Simulation Results}
\label{sec:simulation} 
\begin{figure}[ht]
\centerline{\psfig{file=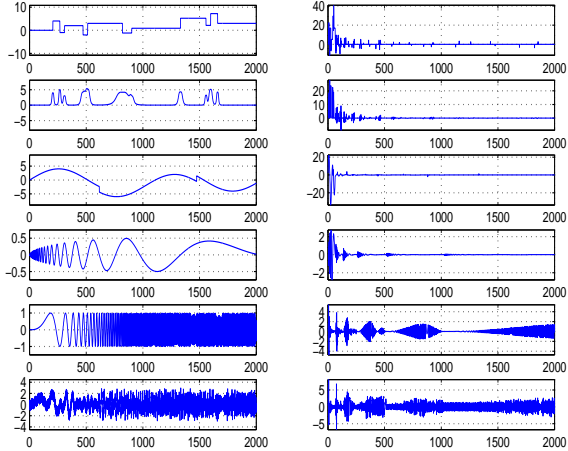,height=7cm,width=9cm}}
\caption{From top to bottom: Blocks, Bumps, HeavySin, Doppler,
QuadChirp, and MishMash. Left figures are the signals and right
figures are their corresponding wavelet coefficients. }
\label{test}
\end{figure}

\begin{figure}[ht]
\centerline{\psfig{file=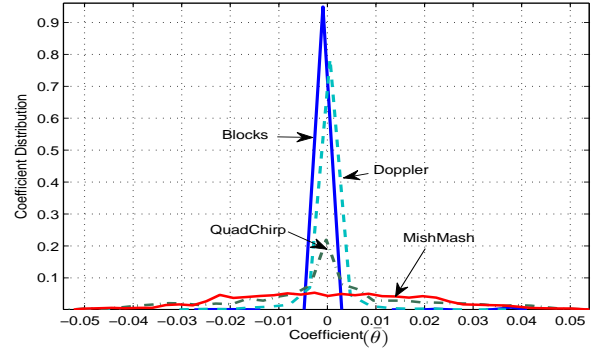,height=5cm,width=9cm}} \caption{
Coefficient distribution for Blocks, Doppler, Quadchirp and
Mishmash. } \label{dist}
\end{figure}

We perform our denoising methods on noisy
versions of six standard signals, Blocks, Mishmash, Bumps, and
Quadchirp which are the test signals introduced in \cite{visu}.
Five level decomposition with Haar wavelet is chosen for this
experiment. The confidence probability of the methods for noise
invalidation region is $0.999997$. Figure \ref{test} shows the six
signals and their coefficients. As this figure confirms, the test
signals represent a wide range of possible coefficient structures.
For example Figure \ref{dist} shows the coefficient distribution
of some of these signals. while signals such as Blocks have very
few nonzero coefficients and many coefficients close to zero,
signals such as mishmash have more uniformly distributed
coefficients. Blocks and MishMash signals represent two extreme
structures, while QuadChirp have a combined structure of both of
these signals.
We compare the proposed method with Visushrink, Sureshrink which
are more general-purpose thresholding approaches. On the other
hand, Sure-LET and BayeshShrink are image denoising methods that
are performing well with the one dimensional signals. We also
consider these method for comparison. We compare the performance of
the methods based on their normalized reconstruction mean square
error (MSE) which is
\begin{eqnarray}
\frac{\|\hat y^N_T-\bar{y}^N\|^2}{||\bar y^N||} \label{nrm}
 \end{eqnarray} where $\hat y^N_T$ (\ref{sahand}) is the
resulted denoised data and $\bar y^N$ is the
noise-free data.
 Table
\ref{tab} provides the MSE of the compared methods. As the table
shows, NIDe performs better than the other approaches in most of
the cases.

\begin{figure}[ht]
\centerline{\psfig{file=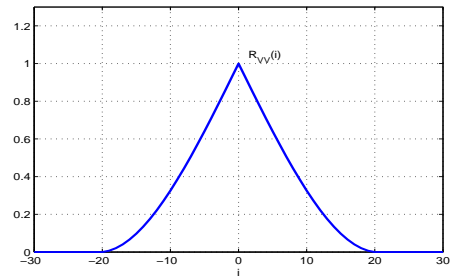,height=4cm,width=7cm}} \caption{
Autocorrelation of the colored additive noise. } \label{dist2}
\end{figure}

The methods are compared for a range of SNRs both for
white or colored noise. Autocorrelation of the  considered
colored noise is shown in Figure \ref{dist2}. The results of
average of 100 runs are provided in Table \ref{tab}. It is
important to note that the MSEs for all
the methods have small standard deviation, much
smaller than the MSE itself. As the table shows, three methods
NIDe, BayesShrink and Sure-Let are comparable in presence of
additive white noise. It is worth mentioning that NIDe outperforms the other two
methods for the more sparse signals $90\%$ of times for even a
wider range of SNR than the range shown in the table. For the non-sparse ones such as Quad-Chirp and
Mishmash, however, Sure-let  performs  slightly better than the
other two methods with additive white noise. For the additive colored noise, NIDe is
consistently outperforming the other methods.

\begin{figure*}[ht!]
  \includegraphics[height=7cm]{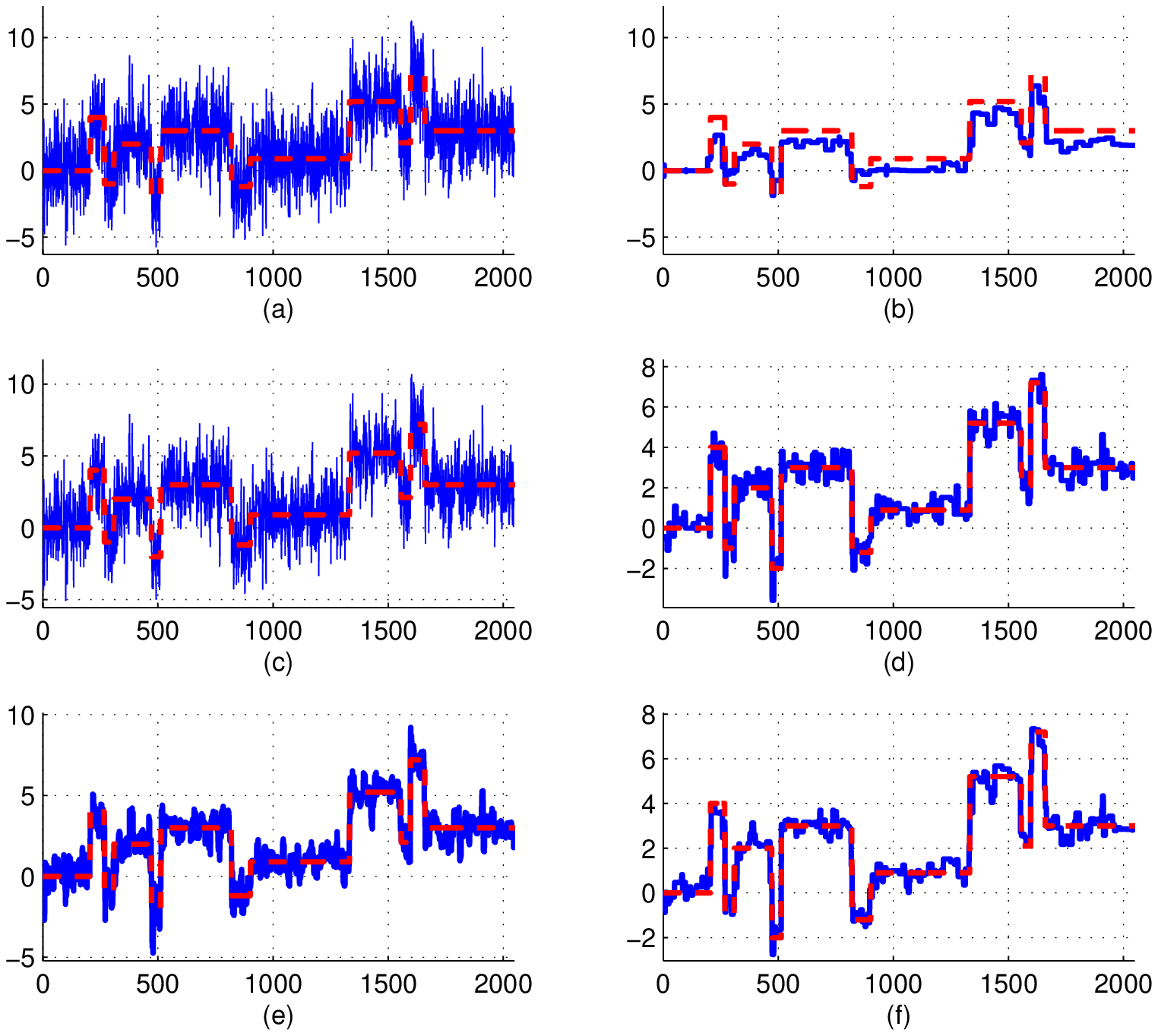}
  \includegraphics[height=7cm]{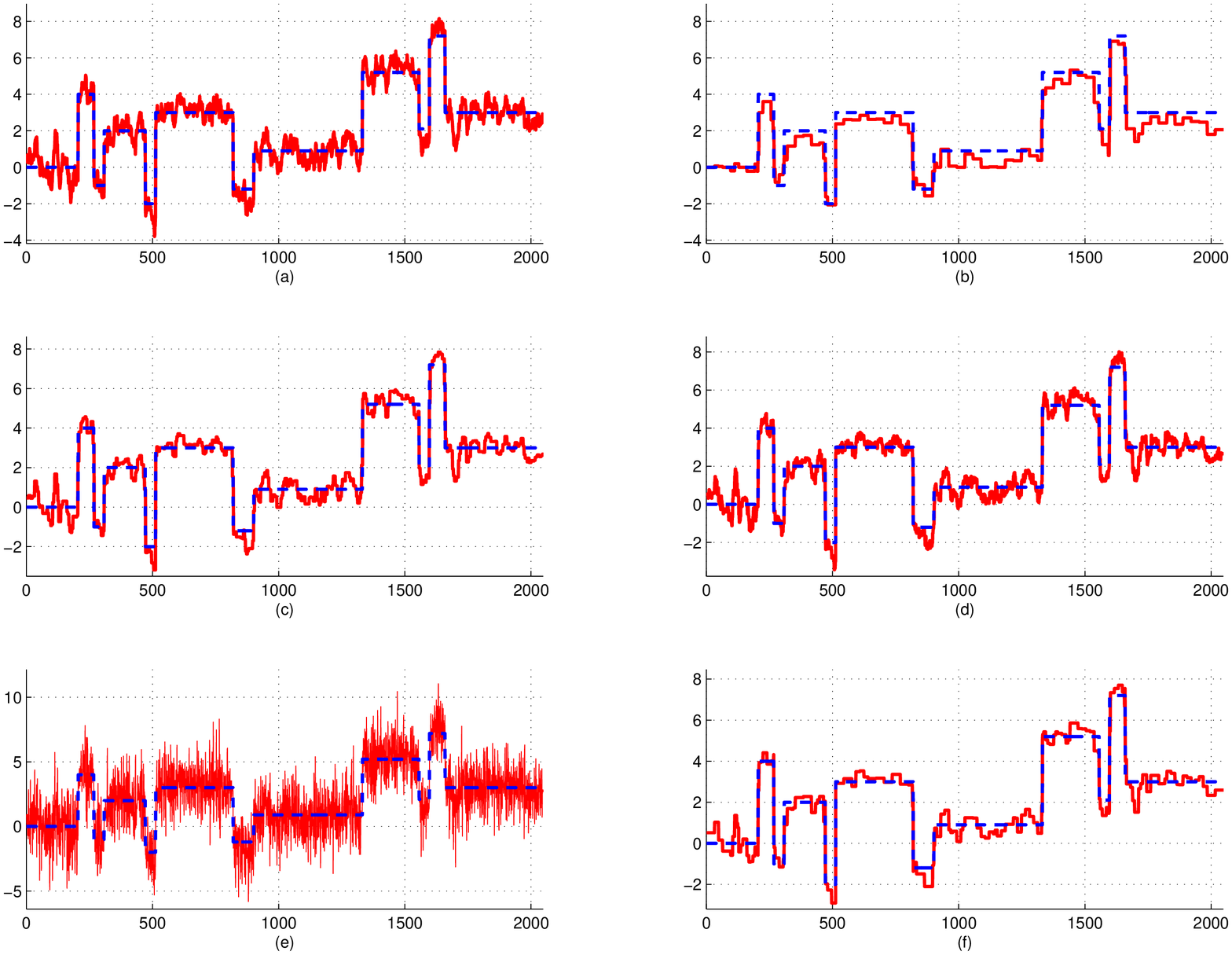}\\
  \caption{Left figure is for additive white noise with SNR=4 and right figure is for additive colored noise with
  SNR=14:  (a) Noisy Blocks, (b) VISU, (c) SURE, (d)
  BayesShrink, (e) Sure-Let, (f) NIDE}\label{color}
\end{figure*}

Figures \ref{color} show the denoised versions of Blocks with
these methods for white and colored additive noise. As the figures
show and Table I confirms, the denoised data with NIDe and
BayesShrink in presence of a white noise are comparable while the
other methods have larger MSE. In presence of colored noise,
however, NIDe method outperforms the other methods.
\begin{table*}[ht!]
\begin{center}
\caption{\label{tab} Normalized Reconstruction MSE for the Thresholding
Methods. Right Table is for the white additive noise and left table is
for the colored additive noise with autocorrelation in Figure \ref{dist2}.
Averaged over 100 runs}
    \begin{tabular}{|c|c|c|c|c|c|}
\hline
     & Visu & SURE & Bayes & SURE & NIDe \\
 &  &  & Shrink & -LET &  \\
\hline
     Blocks&   &    &    &    &    \\
\hline
        SNR=1 & 0.168 & 0.657 & 0.111 & 0.124 & {\bf 0.066} \\ 
        SNR=4 & 0.124 & 0.259 & 0.058 & 0.065 & {\bf 0.042} \\ 
        SNR=8 & 0.086 & 0.032 & 0.026 & 0.027 & {\bf 0.020} \\
        SNR=10& 0.072 & 0.015 &  0.015 & 0.017 & {\bf 0.013} \\
        SNR=14& 0.048 & 0.007 &  0.007 & 0.007 & {\bf 0.005} \\
\hline
        Bumps &   &    &    &    &    \\
\hline
        SNR=1 & 0.155 & 0.427 & 0.098 & 0.122 & {\bf 0.091} \\ 
        SNR=4 & 0.127 & 0.065 & 0.073 & 0.063 & {\bf 0.070} \\ 
        SNR=8 & 0.096 & 0.026 & {\bf 0.023} & 0.027 & 0.025 \\
        SNR=10& 0.083 & 0.023 & 0.019 & 0.018 & {\bf 0.017} \\
        SNR=14& 0.062 & 0.017 & 0.012 & {\bf 0.009} & {\bf 0.009} \\
\hline
        HeavySin &   &    &    &    &    \\
\hline
        SNR=1 & 0.137 & 0.670 & 0.029 & 0.115 & {\bf 0.028} \\ 
        SNR=4 & 0.096 & 0.268 & {\bf 0.017} & 0.057 & {\bf 0.017} \\ 
        SNR=8 & 0.063 & 0.032 & 0.010 & 0.023 & {\bf 0.009} \\
        SNR=10& 0.050 & {\bf 0.007} & 0.016 & 0.015 & \bf 0.007 \\
        SNR=14& 0.035 &  0.004 & {\bf 0.003} & 0.006 & \bf 0.003 \\
\hline
        Doppler &   &    &    &    &    \\
\hline
        SNR=1 & 0.798 & 0.098 & {\bf 0.069} & 0.126 & 0.078 \\ 
        SNR=4 & 0.659 & 0.085 & 0.085 & 0.067 & {\bf 0.076} \\ 
        SNR=8 & 0.493 & 0.078 & 0.036 & 0.030 & {\bf 0.032} \\
        SNR=10& 0.424 & 0.076 & 0.029 & 0.020 & {\bf 0.024} \\
        SNR=14& 0.308 & 0.071 & 0.012 & 0.010 & {\bf 0.009} \\
\hline
        QuadChirp &   &    &    &    &    \\
\hline
        SNR=1 & 0.931 & 0.782 & 0.466 & {\bf 0.447} & 0.637 \\ 
        SNR=4 & 0.903 & 0.771 & 0.284 & {\bf 0.276} & 0.357 \\ 
        SNR=8 & 0.864 & 0.757 & {\bf 0.129} & 0.131 & 0.151 \\
        SNR=10& 0.841 & 0.753 & {\bf 0.086} & \bf 0.086 & 0.096 \\
        SNR=14& 0.780 & 0.750 &  0.038 & 0.037 & \bf 0.035 \\
\hline
        MishMash &   &    &    &    &    \\
\hline
        SNR=1 & 0.926 & 0.523 & 0.517 & {\bf 0.462} & 0.693 \\ 
        SNR=4 & 0.906 & 0.430 & 0.318 & {\bf 0.286} & 0.374 \\ 
        SNR=8 & 0.865 & 0.432 & 0.146 & {\bf 0.136} & 0.156 \\
        SNR=10& 0.836 & 0.623 & 0.095 & {\bf 0.089} & 0.099 \\
        SNR=14& 0.752 & 0.641 & 0.040 & {\bf 0.039} & \bf 0.039 \\
\hline
    \end{tabular}
    \begin{tabular}{|c|c|c|c|c|c|}
\hline
     & Visu & SURE & Bayes & SURE & NIDe \\
 &  &  & Shrink & -LET &  \\
    \hline
     Blocks&   &    &    &    &    \\
\hline
        SNR=1 & 0.406 & 0.695 & 0.562 & 6.212 & {\bf 0.385} \\ 
        SNR=4 & 0.212 & 0.334 & 0.286 & 3.107 & {\bf 0.210} \\ 
        SNR=8 & 0.098 & 0.124 & 0.121 & 1.214 & {\bf 0.092} \\
        SNR=10& 0.063 & 0.074 & 0.079 & 0.753 & {\bf 0.060} \\
        SNR=14& 0.030 & 0.035 & 0.037 & 0.295 & {\bf 0.024} \\
\hline
     {Bumps }&   &    &    &    &    \\
\hline
        SNR=1 & 0.455 & 0.648 & 0.559 & 6.443 &  {\bf 0.427} \\ 
        SNR=4 & 0.279 & 0.305 & 0.293 & 3.220 & {\bf 0.245} \\ 
        SNR=8 & 0.105 & 0.115 & 0.125 & 1.273 & {\bf 0.098} \\
        SNR=10& {\bf 0.072} & 0.075 & 0.081 & 0.773 & {\bf 0.072} \\
        SNR=14&  0.035 & 0.038 & 0.035 & 0.299 & {\bf 0.030} \\
\hline
        {HeavySin} &   &    &    &    &    \\
\hline
        SNR=1 & 0.402 & 0.712 & 0.531 & 6.527 & {\bf 0.334} \\ 
        SNR=4 & 0.209 & 0.328 & 0.270 & 3.256 & {\bf 0.173} \\ 
        SNR=8 & 0.085 & 0.117 & 0.105 & 1.280 & {\bf 0.071} \\
        SNR=10& 0.057 & 0.069 & 0.068 & 0.803 & {\bf 0.047} \\
        SNR=14& 0.027 & 0.023 & 0.028 & 0.317 & {\bf 0.020} \\
\hline
Doppler &   &    &    &    &    \\
\hline
        SNR=1 & 0.446 & {\bf 0.399} & 0.569 & 6.431 & 0.401 \\ 
        SNR=4 & 0.330 & 0.236 & 0.293 & 3.201 & {\bf 0.235} \\ 
        SNR=8 & 0.231 & 0.138 & 0.127 & 1.207 & {\bf 0.110} \\
        SNR=10& 0.187 & 0.113 & 0.082 & 0.76 & {\bf 0.075} \\
        SNR=14& 0.133 & 0.089& 0.037  & 0.038  & {\bf 0.030} \\
\hline
        QuadChirp &   &    &    &    &    \\
\hline
        SNR=1 & 0.964 & 1.183 & 1.139 & 1.298 & {\bf 0.594} \\ 
        SNR=4 & 0.978 & 0.965 & 0.489 & 0.536 & {\bf 0.326} \\ 
        SNR=8 & 0.986 & 0.833 & 0.164 & 0.186 & {\bf 0.137} \\
        SNR=10& 0.990 & 0.801 & 0.100 & 0.117 & {\bf 0.088} \\
        SNR=14& 0.980 & 0.771 & 0.039 & 0.051 & {\bf 0.037} \\
\hline
        MishMash &   &    &    &    &    \\
\hline
        SNR=1 & 0.974 & 1.186 & 1.205 & 1.246 & {\bf 0.607} \\ 
        SNR=4 & 0.956 & 0.889 & 0.523 & 0.523 & {\bf 0.336} \\ 
        SNR=8 & 0.961 & 0.727 & 0.174 & 0.180 & {\bf 0.142} \\
        SNR=10& 0.963 & 0.699 & 0.105 & 0.109 & {\bf 0.091} \\
        SNR=14& 0.963 & 0.675 & 0.042 & 0.041 & {\bf 0.031} \\
\hline
    \end{tabular}
\end{center}
\end{table*}

\section{Conclusion}\label{con}
A denoising approach based on direct invalidation of the
coefficients is proposed. The invalidation process uses a
signature of the additive noise in the form of a probabilistic
confidence region. The signature is defined based on the
statistical properties of the additive noise and is such that its
standard deviation is much smaller than its mean. In this work we
provided one example of such signature which illustrates itself in
simply sorting the coefficients. It was shown that such a
signature represents the noise in a dense area. The density of the
area depends on the noise variance and the data length. The
smaller is the noise variance and/or the longer is the data
length, the denser is the signature area. This will enable us
to invalidate whether every coefficient is noise dominant or
 data dominant. The theory of the method shows its strength
for any type of noise-free signal. The variance of the
signature decreases as the data length grows, providing more
distinguish and denser area. The method denoises in
presence of not only an additive white noise, but also an additive colored noise 
This will also enable the user to use the proposed
 thresholding technique with any non-orthogonal basis. Simulation results confirmed the advantages of
the proposed noise invalidation approaches in terms of the
reconstruction MSE and illustrates its robust
performance. 
 While the proposed NIDe approach is a pioneer
method for a general-purpose thresholding, there seems to be a
great potential in further analysis, study, and expansion of such
invalidation method for the purpose of denoising. For example,
it is worth investigating further functions of different noise
distributions that can serve as the noise signature.
In addition, for the cases that the noise-free signal
belongs to a particular class of signals, a potential extension
would be to combine the statistical properties of the noise-free
signal with the statistical structure of the additive noise in
order to define noise signatures for the purpose of
probabilistic invalidation.

\appendices
\section{Mean and Variance of the Signature for IID noise}\label{mv}
For the expected value of this signature from (\ref{onne2}) we have
\begin{eqnarray}
E(g(z,V^N))=\frac{1}{N}\sum_{i=1}^NE(g(z,V_i))
=G_E(z)
\end{eqnarray}
For the variance of the signature, since $V_i$ and $V_j$ are independent we have
\begin{eqnarray}
E(g(z,V_i)g(z,V_j))=E(g(z,V_i)E(g(z,V_j))
\end{eqnarray}
Therefore, the cross terms in the desired variance are zero and from (\ref{for4}) we have
\begin{eqnarray}
var(g(z,V^N))=\frac{1}{N^2}\sum_{i=1}^Nvar(g(z,V_i))
=\frac{1}{N}G_{var}(z)
\end{eqnarray}
\section{Mean and Variance of the Sorting Signature for IID Noise}\label{gol}
For the mean we have
\begin{eqnarray}
E(g(z,V_i))=1\times Pr(|V_i|\leq z)+0\times Pr(|V_i|>z)\nonumber\\
=Pr(V_i\leq z)=F(z) \label{onne}
\end{eqnarray}
Therefore,
\begin{eqnarray}
E(g(z,V^N))=\frac{1}{N}\sum_{i=1}^NE(g(z,V_i))
=F(z)
\end{eqnarray}
For the variance
\begin{eqnarray}
E(g^2(z,V_i)=1\times Pr(|V_i|\leq z)+0\times Pr(|V_i|>z)\nonumber\\
=Pr(|V_i|\leq z)=F(z) \label{twwo}
\end{eqnarray}
From (\ref{onne}) and (\ref{twwo})
\begin{eqnarray}
var(g(z,V_i)=E(g^2(z,V_i))-(E(g(z,V)))^2\nonumber\\
=F(z)-F^2(z)\label{for}
\end{eqnarray}
On the other hand
\begin{eqnarray}
E(g(z,V_i)g(z,V_j))=1\times Pr(|V_i|\leq z \;\&\; |V_j|\leq z)+\nonumber
\\0\times Pr(|V_i|>z \;or\; |V_j|>z)=\nonumber\\
Pr(V_i\leq z)\Pr(|V_j|\leq z)=F^2(z)
\end{eqnarray}
Therefore, we have
\begin{eqnarray}
E((g(z,V_i)-F(z))(g(z,V_j)-F(z)))=0 \label{tree}
\end{eqnarray}
which sets the cross terms in the variance of $g(z,V)$ to zero. From (\ref{for}), (\ref{tree}) we have
\begin{eqnarray}
var(g(z,V^N))=\frac{1}{N^2}\sum_{i=1}^Nvar(g(z,V_i))\nonumber\\
=\frac{1}{N}F(z)(1-F(z))
\end{eqnarray}
\section{Mean and Variance of the Data with Sorting Signature}\label{golnar}
 \begin{eqnarray}
E(g(z,\theta_i))=Pr(|v_i+\bar\theta_i| \leq z)\nonumber\\
Pr(-z-\bar\theta_i\leq v_i \leq z-\bar\theta_i)=\phi(\frac{z-\bar\theta_i}{\sigma})-\phi(\frac{-z-\bar\theta_i}{\sigma})
\end{eqnarray}
since $\phi(-a)=1-\phi(a)$, we have
\begin{eqnarray}
-\phi(\frac{-z-\bar\theta_i}{\sigma})=\phi(\frac{z+\bar\theta_i}{\sigma})-1
\end{eqnarray}
which results $E(g(z,\theta_i))$ to be the defined $H(z,\bar\theta_i)$ in (\ref{H}). For the variance, due to the structure of $g(z,\bar\theta_i)$, similar to what we had for the noise, we have
 \begin{eqnarray}
E(g^2(z,\theta_i))=E(g(z,\theta_i))=H(z,\bar\theta_i) \label{power}
\end{eqnarray}
Therefore
\begin{eqnarray}
var(g(z,\theta_i))=E(g^2(z,\theta_i))-(E(g(z,\theta_i)))^2\nonumber\\
=H(z,\bar\theta_i)(1-H(z,\bar\theta_i))
\end{eqnarray}
Which results in the variance in (\ref{vartheta}).
\section{Mean and Variance of the Signature for Colored Noise}\label{gol2}
\begin{figure}[ht]
\centerline{\psfig{file=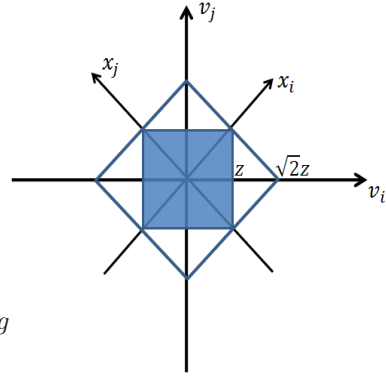,height=5cm,width=5cm}} \caption{
The desired area for calculation of the probabilities in (\ref{bnd1}) and (\ref{pp}).} \label{boundaries}
\end{figure}

In this case the autocorrelation between the zero mean Gaussian $v_i$s is denoted by $R_{vv}(m)$.
\begin{eqnarray}
g(z,v^N))=\frac{1}{N}\sum_{i=1}^Ng(z,v_i)=[\frac{1}{N}\; \frac{1}{N}\;\cdots \;\frac{1}{N}]\left[\begin{array}{c} g(z,v_1) \\ g(z,v_2)\\ \vdots\\
g(z,v_N) \end{array} \right]
\end{eqnarray}
For the expected value of this function we have
\begin{eqnarray}
E(g(z,V^N)))=\frac{1}{N}\sum_{i=1}^NE(g(z,V_i))=F(z)
\end{eqnarray}
which is similar to that of the IID additive noise. However, for the variance, since the following holds
 \begin{eqnarray}
var(g(z,V^N))=var(\frac{1}{N}\sum_{i=1}^Ng(z,V_i))=\nonumber
\end{eqnarray}
\begin{eqnarray}
[\frac{1}{N}\; \frac{1}{N}\;\cdots \;\frac{1}{N}]\left[\begin{array}{cccc} var(g(z,V_1)) & cov(g(z,V_1)g(z,V_2) &\cdots & cov(g(z,V_1)g(z,V_N)) \\ cov(g(z,V_1)g(z,V_2))& var(g(z,V_2)) & \cdots & \vdots\\ \vdots & \vdots & \ddots &\vdots \\ cov(g(z,V_1)g(z,V_N)) & \cdots & \cdots &  var(g(z,V_N)))
 \end{array} \right] \left[\begin{array}{c} \frac{1}{N} \\ \frac{1}{N}\\ \vdots\\ \frac{1}{N}\\ \end{array} \right]\nonumber
\end{eqnarray}
 we have
\begin{eqnarray}
var(g(z,V^N)))=\frac{1}{N}F(z)(1-F(z))+ \nonumber \\ \frac{1}{N^2}\sum_{i=1,j=1,i\neq j}^N cov(g(z,V_i)g(z,V_j)) \label{varr}
\end{eqnarray}
where elements of the second term are
\begin{eqnarray}
cov(g(z,V_i)g(z,V_j))= \;\;\;\;\;\;\;\;\;\;\;\;\;\;\;\;\;\;\;\;\;\;\;\;\;\nonumber \\E(g(z,V_i)(g(z,V_i))- E(g(z,V_i))E(g(z,V_i))
\end{eqnarray}
where the second term is simply $F^2(z)$, the first term is $Pr(|V_i|\leq z \;\&\; |V_j|\leq z)$.
For the first term,
the joint distribution of $V_i$ and $V_j$ is a Gaussian distribution with zero mean and variance
\begin{eqnarray}
E(\left[\begin{array}{c}  V_i\\ V_j\\
\end{array} \right])=\left[\begin{array}{c}  0\\ 0\\
\end{array} \right],\;
var(\left[\begin{array}{c}  V_i\\ V_j\\
\end{array} \right])=\sigma^2\left[\begin{array}{cc} 1 & \rho_{ij}\\ \rho_{ij} & 1\\
\end{array} \right]
\end{eqnarray}
where
\begin{eqnarray}
\rho_{ij}=\frac{R_{vv}(i-j)}{R_{vv}(0)}
\end{eqnarray}
with $R_{vv}(0)=\sigma^2$.
The decomposition of the covariance matrix is as follows
\begin{eqnarray}
\left[\begin{array}{cc} 1 & \rho_{ij}\\ \rho_{ij} & 1\\
\end{array} \right]= \;\;\;\;\;\;\;\;\;\;\;\;\;\;\;\;\;\;\;\;\;\;\;\;\;\; \nonumber \\ \frac{1}{2}\left[\begin{array}{cc} 1 & 1\\ 1 & -1 \\
\end{array} \right]\left[\begin{array}{cc} 1+\rho_{ij} & 0\\ 0 & 1-\rho_{ij} \\
\end{array} \right]\left[\begin{array}{cc} 1 & 1\\ 1 & -1 \\
\end{array} \right]
\end{eqnarray}
Therefore, by the following transformation
\begin{eqnarray}
\left[\begin{array}{c} x_i\\ x_j
\end{array} \right]= \frac{1}{\sqrt{2}}\left[\begin{array}{cc} 1 & 1\\ 1 & -1 \\
\end{array} \right]\left[\begin{array}{c} v_i\\ v_j \\
\end{array} \right]
\end{eqnarray}
the two $x_i$ and $x_j$ random variables are independent and with variances $\sigma^2(1+\rho_{ij})$  and $\sigma^2(1-\rho_{ij})$. As a result, for the first term of the covariance
\begin{eqnarray}
E(g(z,V_i)(g(z,V_i))=Pr(|V_i|\leq z \;\&\; |V_j|\leq z) \label{bnd1} \\
 \leq Pr(|X_i|\leq \sqrt{2}z \;\&\; |X_j|\leq \sqrt{2}z)\\=F(\frac{\sqrt{2}z}{\sqrt{1+\rho_{ij}}})F(\frac{\sqrt{2}z}{\sqrt{1-\rho_{ij}}})\label{pp}
\end{eqnarray}
Figure \ref{boundaries} show the area considered for the calculation of this probability.
\subsection{Noisy Data}
With similar analogy, for the signal in presence of the noisy data, we have
\begin{eqnarray}
E(g(z,\Theta^N)))=\frac{1}{N}\sum_{i=1}^N H(z,\bar\theta_i)\label{ahh}
\end{eqnarray}
where $H(z,\bar\theta_i)$ was defined in (\ref{H}). For the variance of the sorted absolute value of the noisy data, similarly (\ref{power}) holds. Therefore, structure of the variance is similar to (\ref{varr}):
 \begin{eqnarray}
var(g(z,\Theta^N)))=\frac{1}{N^2}\sum_{i=1}^N (H(z,\bar\theta_i)-H^2(z,\bar\theta_i))+ \nonumber \\ \frac{1}{N^2}\sum_{i=1,j=1,i\neq j}^N cov(g(z,\theta_i)g(z,\theta_j))
\end{eqnarray}
For the covariance in the second term
\begin{eqnarray}
cov(g(z,\Theta_i)g(z,\Theta_j))= \;\;\;\;\;\;\;\;\;\;\;\;\;\;\;\;\;\;\;\;\;\;\;\;\;\nonumber \\E(g(z,\Theta_i)(g(z,\Theta_i))- E(g(z,\Theta_i))E(g(z,\Theta_i))
\end{eqnarray}
we use (\ref{ahh}) to calculate $E(g(z,\Theta_i))E((g(z,\Theta_i))$
and have
\begin{eqnarray}
E(g(z,\Theta_i)(g(z,\Theta_i))=Pr(|V_i+\bar\theta_i|\leq z \;\&\; |V_j+\bar\theta_j|\leq z) \nonumber \\
 \leq Pr(\sqrt{2}(-z-\frac{\bar\theta_i+\bar\theta_j}{2})\leq X_i\leq \sqrt{2}(z-\frac{\bar\theta_i+\bar\theta_j}{2}) \\ \;\&\; \sqrt{2}(-z-\frac{\bar\theta_i-\bar\theta_j}{2})\leq X_j\leq \sqrt{2}(z-\frac{\bar\theta_i-\bar\theta_j}{2})) \\=H(\frac{\sqrt{2}z}{\sqrt{1+\rho_{ij}}},\frac{\bar\theta_i+\bar\theta_j}{\sqrt{2(1+\rho_{ij})}})
 H(\frac{\sqrt{2}z}{\sqrt{1-\rho_{ij}}},\frac{\bar\theta_i-\bar\theta_j}{\sqrt{2(1-\rho_{ij})}})
\end{eqnarray}

\end{document}